\documentclass[twocolumn,showpacs,preprintnumbers,amsmath,amssymb]{revtex4}

\usepackage{graphicx}
\usepackage{dcolumn}
\usepackage{bm}

\begin{document}

\preprint{APS/123-QED}

\title{Finite size scaling and equation of state for Ising lattices}

\author{J.G. Garc\'{\i}a}
 \email{jg.garcia@uam.es}
\author{J.A. Gonzalo}%
 \email{julio.gonzalo@uam.es}
\affiliation{
Departamento de F\'\i sica de Materiales C-IV,\\
Universidad Aut\'onoma de Madrid.\\
Cantoblanco, 28049 Madrid, Spain.}


\date{\today}

\begin{abstract}

Accurate Monte Carlo data from a set of isotherms near the critical point are
analyzed using two RG based complementary representations, given respectively in
terms of $\bar h$=$h$/$\mid$$t$$\mid^{\beta\delta}$ and $\bar \tau$=$t/h^{1/\beta\delta}$.
Scaling plots for data on simple cubic Ising lattices are compared with plots of
$ML^{\beta\nu}$ vs. $\mid$$t$$\mid$$L^{1/\nu}$ for increasing $L$ values and
with high quality experimental data on $CrBr_{3}$. Finite size effects and the 
equation of state are discussed.

\end{abstract}


\maketitle


Monte Carlo methods \cite{Binder} using Wolff algorithms \cite{Wolff} have been
extensively used to
describe $M(T)$ for $H$=0 in the whole range of temperatures both below
and above the critical temperature (specially at $T$$\cong$$T_c$) but few,
if any, Monte Carlo simulations of magnetic isotherms $M(H)$ at $T$$\cong$$T_c$
have been reported in the literature. It is clear, however, that with recent
improvements in computing facilities (larger memory, greater speed, better
availability) accurate, well thermalized, closely spaced data can provide
very substantial contribution to describing the phase transition and better
understanding of finite size effects.

Simulations of this type are reported in this work, performed using
Metropolis algorithms \cite{Metropolis}, which are specially convenient to describe
the system evolution at constant temperature and for small field
increments\cite{Garcia,Garcia2}. Using relatively large lattices with 70$\times$70$\times$70 spins,
an accurate characterization of the scaling behavior, and, therefore, the
equation of state in the vicinity of the critical point
($H$=0, $T_c$=4.511523785) \cite{Blote} for simple cubic Ising
lattices can be obtained. Details of the Monte Carlo calculations for isotherms
taken at $T$ near $T_c$ are given in reference \cite{Garcia}. Periodic
boundary conditions were used and 140,000 Monte Carlo steps were
taken at each field/temperature to ensure equilibrium.

For the scaling representation of the raw $M(H)$ data at each $T$ we
did use two complementary ways: (a) the usual way \cite{Yeomans}, involving
$\bar m$$\equiv$$M(t,h)$/$\mid$$t$$\mid^\beta$ and $\bar h$$\equiv$$h$/$\mid$$t$$\mid^{\beta\delta}$,
where $t\equiv(T-T_c)/T_c$ involves the temperature gap 
to $T_c$, $h$ (reduced field) is proportional to $H$,
$\beta$ is the spontaneous magnetization critical exponent
$\beta\equiv\partial$log$M/\partial$log$\mid$$t$$\mid$ at $T$$\rightarrow$$T_c$, and $\delta$ is the
critical isotherm exponent $\delta^{-1}$$\equiv\partial$log$M/\partial$log$h$ at $T$$\rightarrow$$T_c$;
and (b) a complementary way
involving $\bar \mu$$\equiv$$M(h,t)/h^{1/\delta}$ and $\bar \tau$$\equiv$$t/h^{1/\beta\delta}$.

Near the fixed point corresponding to the critical point, the singular part
of the reduced free energy per spin may be written \cite{Yeomans} in scaling form as

\begin{equation}
\label{eq:eq1}
\bar f_s(g_{1}, g_{2}, g_{3}, \dots) \sim \bar b^{d} \bar f_{s}(b^{y_{1}}g_{1}, b^{y_{2}}g_{2}, b^{y_{3}}g_{3}, \dots)
\end{equation}

where $b$ is an arbitrary scale factor and the $y_{i}$ relate to the
usual critical exponents. To get (a) Equation (\ref{eq:eq1}) is
differentiated with respect to the field in the usual way to obtain

\begin{equation}
\label{eq:eq2}
M(t, h, g_{3}, \dots) \sim \bar b^{-d+y_{2}} M(b^{y_{1}}t, b^{y_{2}}h, b^{y_{3}}g_{3}, \dots)
\end{equation}

Taking $b^{y_{1}}$$\mid$$t$$\mid$=1 and setting the irrelevant variables equal to zero

\begin{equation}
\label{eq:eq3}
M(t, h) \sim \mid t \mid^{(d-y_{2})/y_{1}} M(\pm 1, h \mid t \mid^{-y_{2}/y_{1}})
\end{equation}

and using the well known scaling relationships \cite{Yeomans} giving $y_1$
and $y_2$ in terms of $\beta$ and $\delta$ one gets

\begin{equation}
\label{eq:eq4}
M(t, h) \sim \mid t \mid^{\beta} M(\pm 1, h \mid t \mid^{-\beta\delta})
\end{equation}

which implies scaling using the ordinary scaling variables

\begin{equation}
\label{eq:eq5}
\bar m \equiv M(t, h) / \mid t \mid^{\beta},    \bar h \equiv h / \mid t \mid^{\beta\delta}
\end{equation}

To get (b), on the other hand, we can argue in an analogous way. Taking $b^{y_2}h$=1
in Equation (\ref{eq:eq2}) and setting the irrelevant variables equal to zero,

\begin{equation}
\label{eq:eq6}
M(t, h) \sim h^{(d-y_2)/y_2} M(t\cdot h^{-y_{1}/y_{2}},1)
\end{equation}

and using $y_1$ and $y_2$ in terms of $\beta$ and $\delta$, one finally gets

\begin{equation}
\label{eq:eq7}
M(t, h) \sim h^{1/\delta} M(t^{-1/\beta\delta},1)
\end{equation}

which implies as alternative scaling variables

\begin{equation}
\label{eq:eq8}
\bar \mu \equiv M(t, h) / h^{1/\delta},    \bar \tau \equiv t / h^{1/\beta\delta}
\end{equation}

%
%
\begin{figure}
\includegraphics[width=6.1cm,height=7.9cm,angle=270]{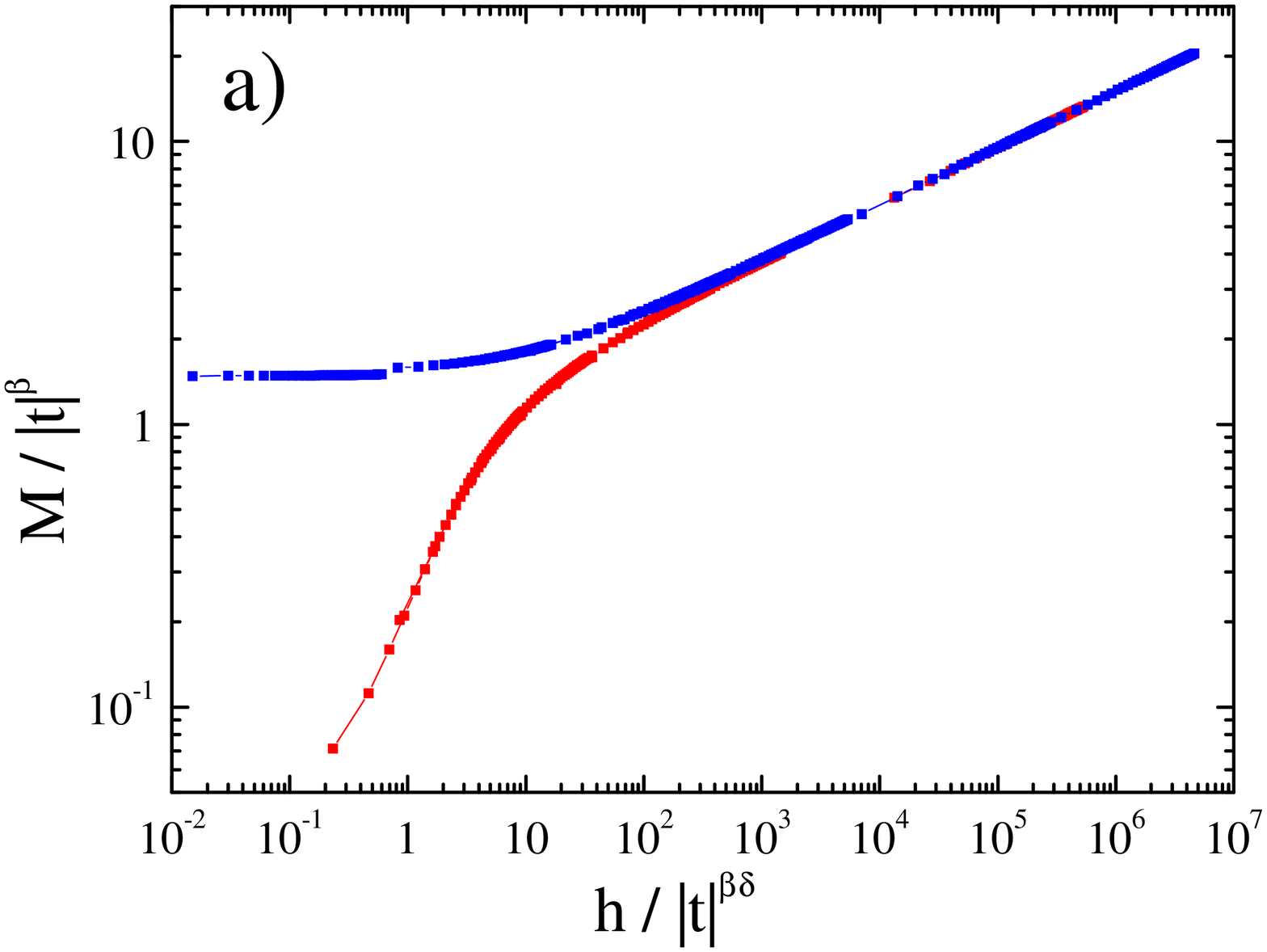}
\\
\includegraphics[width=6.1cm,height=7.9cm,angle=270]{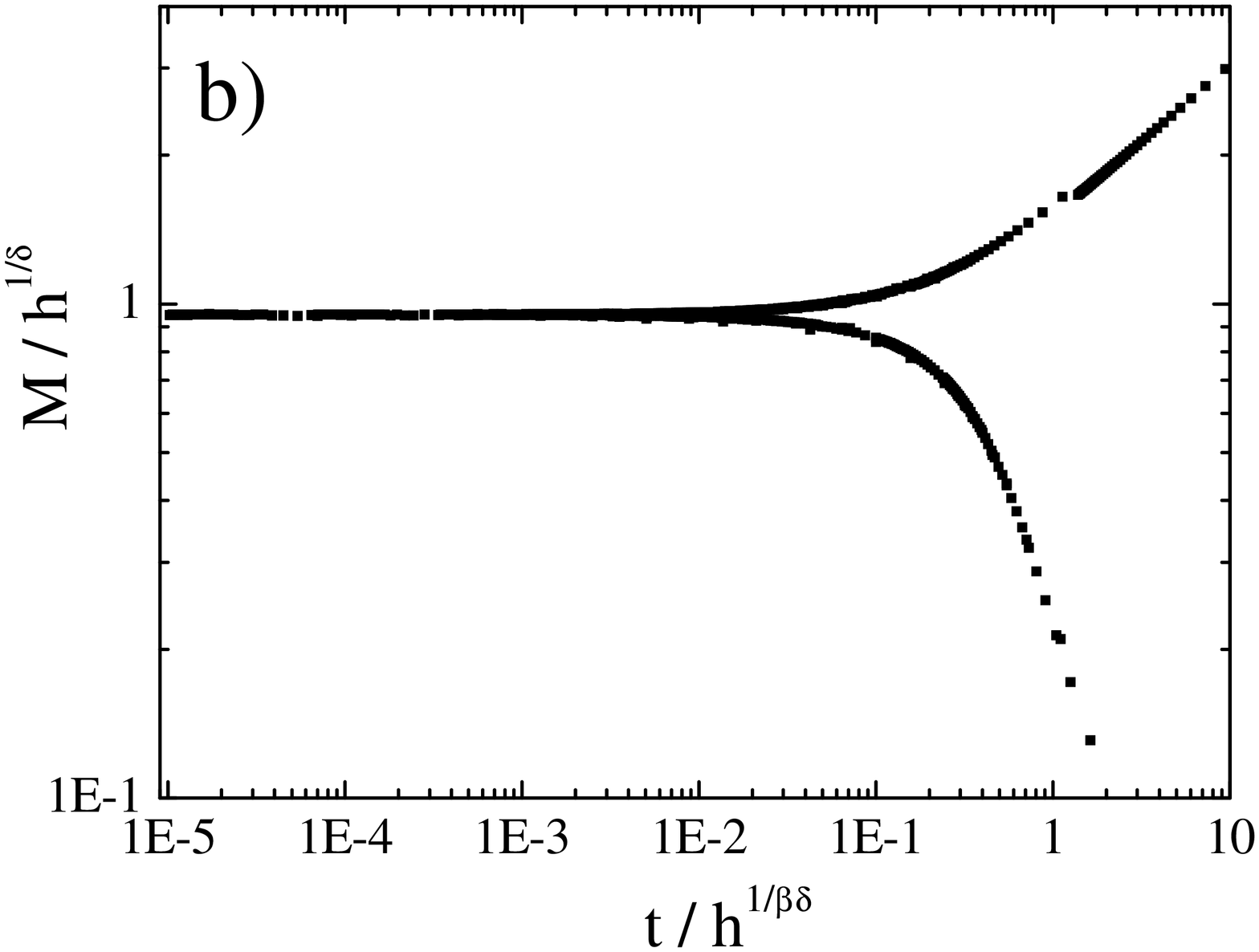}
\caption{\label{fig:epsart1}
Scaling plots of Monte Carlo data for a s.c. Ising lattice
of $70^3$ spins. (a) $\bar m$$\equiv$$M$/$\mid$$t$$\mid^{\beta}$ vs
$\bar h$$\equiv$$h$/$\mid$$t$$\mid^{\beta\delta}$, and
(b) $\bar \mu$$\equiv$$M$/$h^{1/\delta}$ vs $\bar \tau$$\equiv$$t$/$h^{1/\beta\delta}$.
The data include 30 isotherms in the intervals 4 $< T <$ 5
and 0 $< h <$ 0.02, the same symbol has been used for all of them. 
The critical temperature \cite{Blote} was taken as
$T_c$=4.511523785 and the critical exponents \cite{Garcia} as $\beta$=5/16 and $\delta$=5.
}
\end{figure}
%
%

Figure 1(a) gives our Monte Carlo data using $\bar m$ and $\bar h$ as
scaling variables as given by Equation (\ref{eq:eq5}) with $T_c$=4.511523785 and
the exponents $\beta$=5/16=0.3125 and $\delta$=5. The usual behavior is observed.
We may note that the data scale extremely well and that only very minor deviations
at lower $h$, attributable to finite size effects are perceptible.
Figure 1(b) gives the same data using the alternative scaling representation. It
may be noted immediately that the plot in Figure 1(b) resembles closely \cite{Binder}
typical scaling plots of $ML^{\beta/\nu}$ vs $\mid$$t$$\mid$$L^{1/\nu}$ for lattices
with linear dimension $L$, in our case, for Ising systems of $L^3$ spins,
suggesting \cite{Marques} formal relationships between $H$ and $L$, $M$ and $L$,
and $\mid$$t$$\mid$ and $L$, which, through the scaling plot branches defining
the critical isotherm, the spontaneous magnetization (coexistence curve) below $T_c$,
and the low field susceptibility above, imply respectively

\begin{equation}
\label{eq:eq9}
H \sim L^{-\beta\delta/\nu},  M \sim L^{-\beta/\nu}  and  \mid t \mid \sim L^{-1/\nu}
\end{equation}

Hence

$ML^{\beta/\nu}\sim$const  ($t$$\gtrless$0) $\longrightarrow$ critical isotherm

$ML^{\beta/\nu}\sim$const$\times$($\mid$$t$$\mid$$L^{1/\nu}$$)^{\beta}$  ($t<0$) $\longrightarrow$ spontaneous magnetization

$ML^{\beta/\nu}\sim$const$\times$($\mid$$t$$\mid$$L^{1/\nu}$$)^{\beta(\delta-1)/2}$  ($t>0$) $\longrightarrow$ susceptibility

%
%
\begin{figure}
\includegraphics[width=6.1cm,height=7.9cm,angle=270]{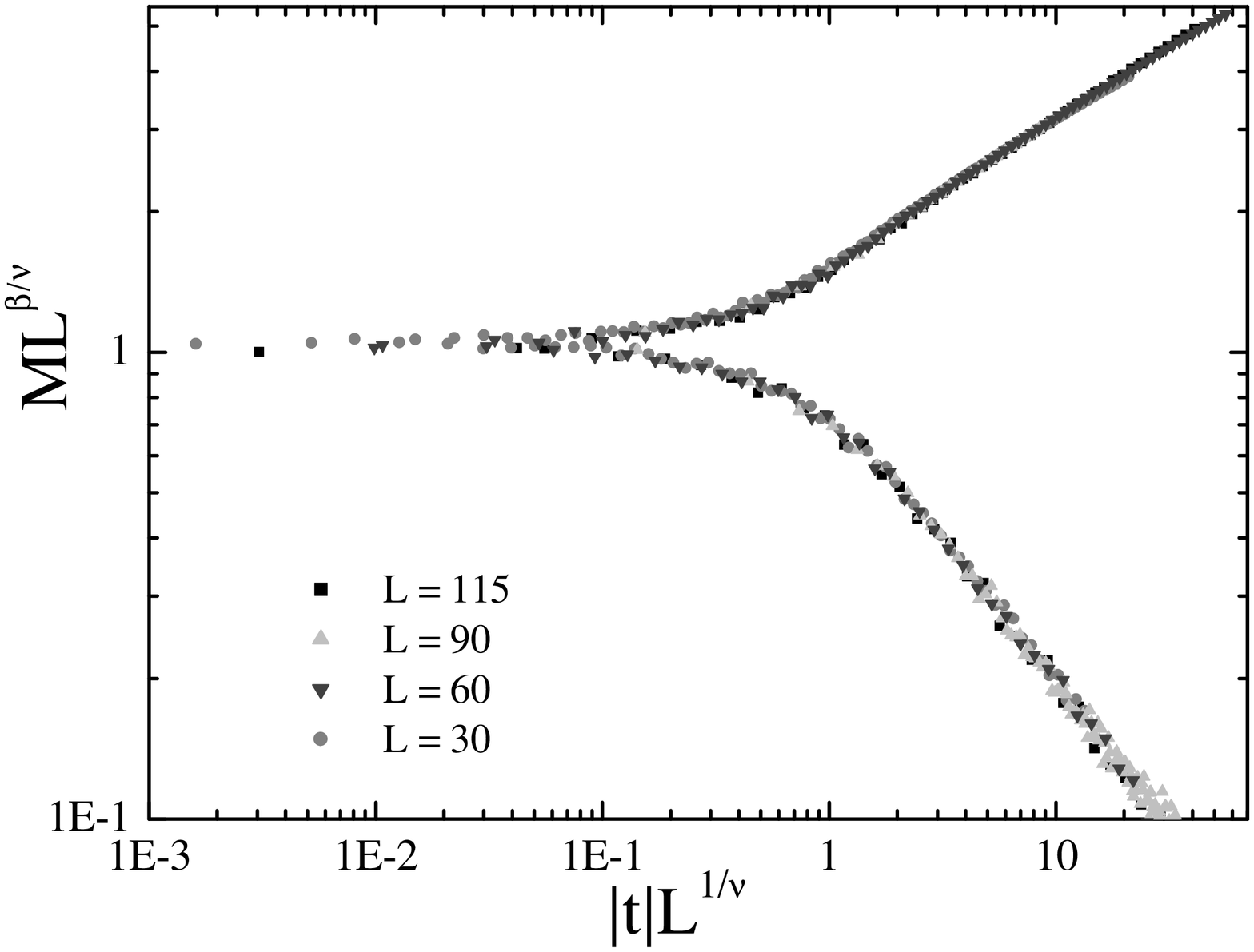}
\caption{\label{fig:epsart2}
Scaling plot of Monte Carlo data $ML^{\beta/\nu}$ vs $\mid$$t$$\mid$$L^{1/\nu}$ for
s.c. Ising lattices with linear size $L$=30, 60, 90, 115. Note that finite size
effects for $L$=30 show up closer to $\mid$$t$$\mid$$\rightarrow$0.}
\end{figure}
%
%

Figure \ref{fig:epsart2} shows data of $ML^{\beta/\nu}$ vs $\mid$$t$$\mid$$L^{1/\nu}$ for
$H$=0 and $L$=30, 60, 90, 115 which mimic the behavior shown in Figure 1(b)
implying that simulations using periodic boundary conditions of phase transitions
with finite size show the effects of an effective straining contribution to
the magnetic field $H_{fs}\sim$$L^{-\beta\delta/\nu}$, i.e. $H_{eff}$=$H$+$H_{fs}$
such that for $L$$\rightarrow$$\infty$, $H_{eff}$$\cong$$H$.

%
%
\begin{figure}
\includegraphics[width=6.1cm,height=7.9cm,angle=270]{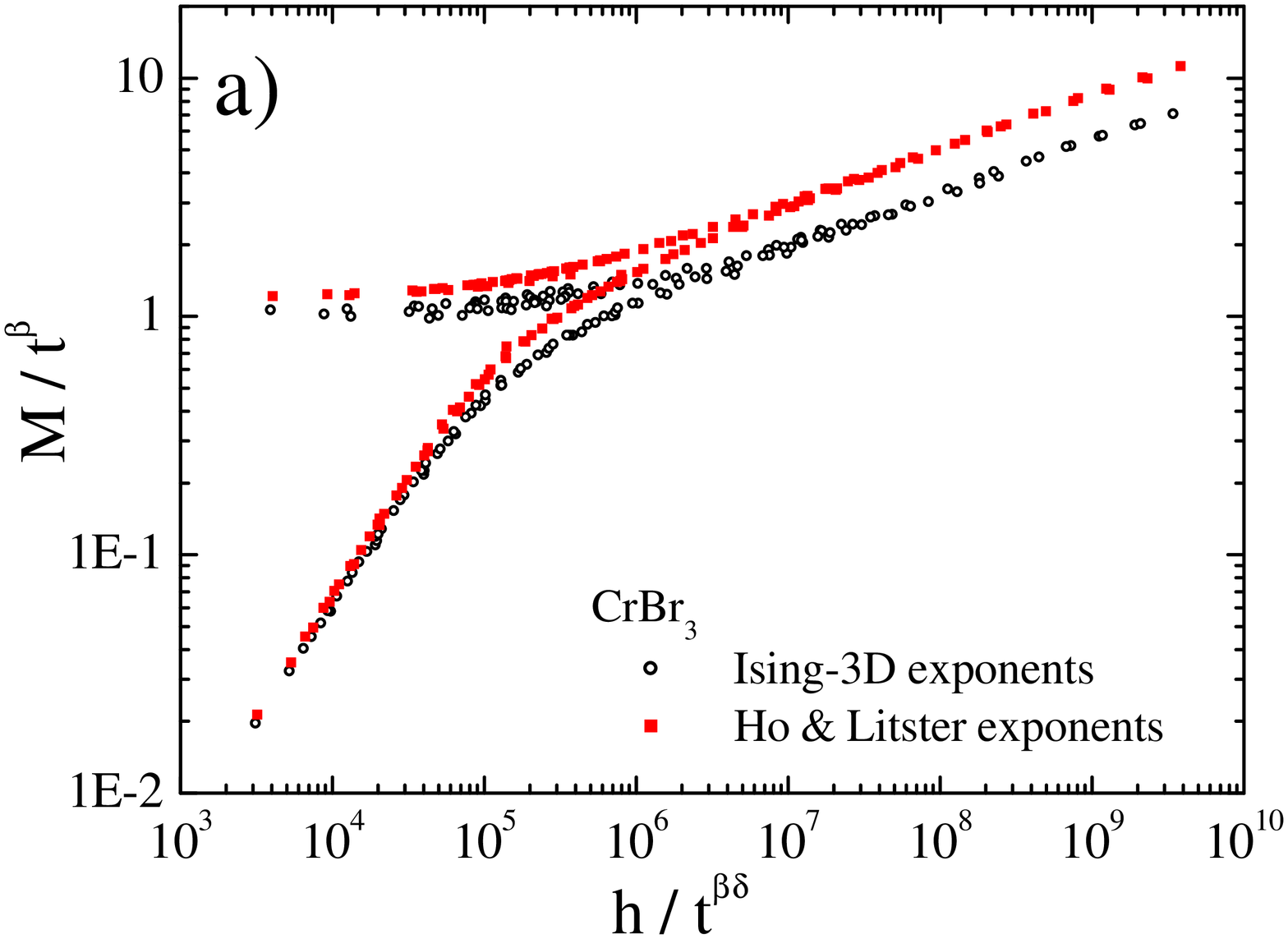}
\\
\includegraphics[width=6.1cm,height=7.9cm,angle=270]{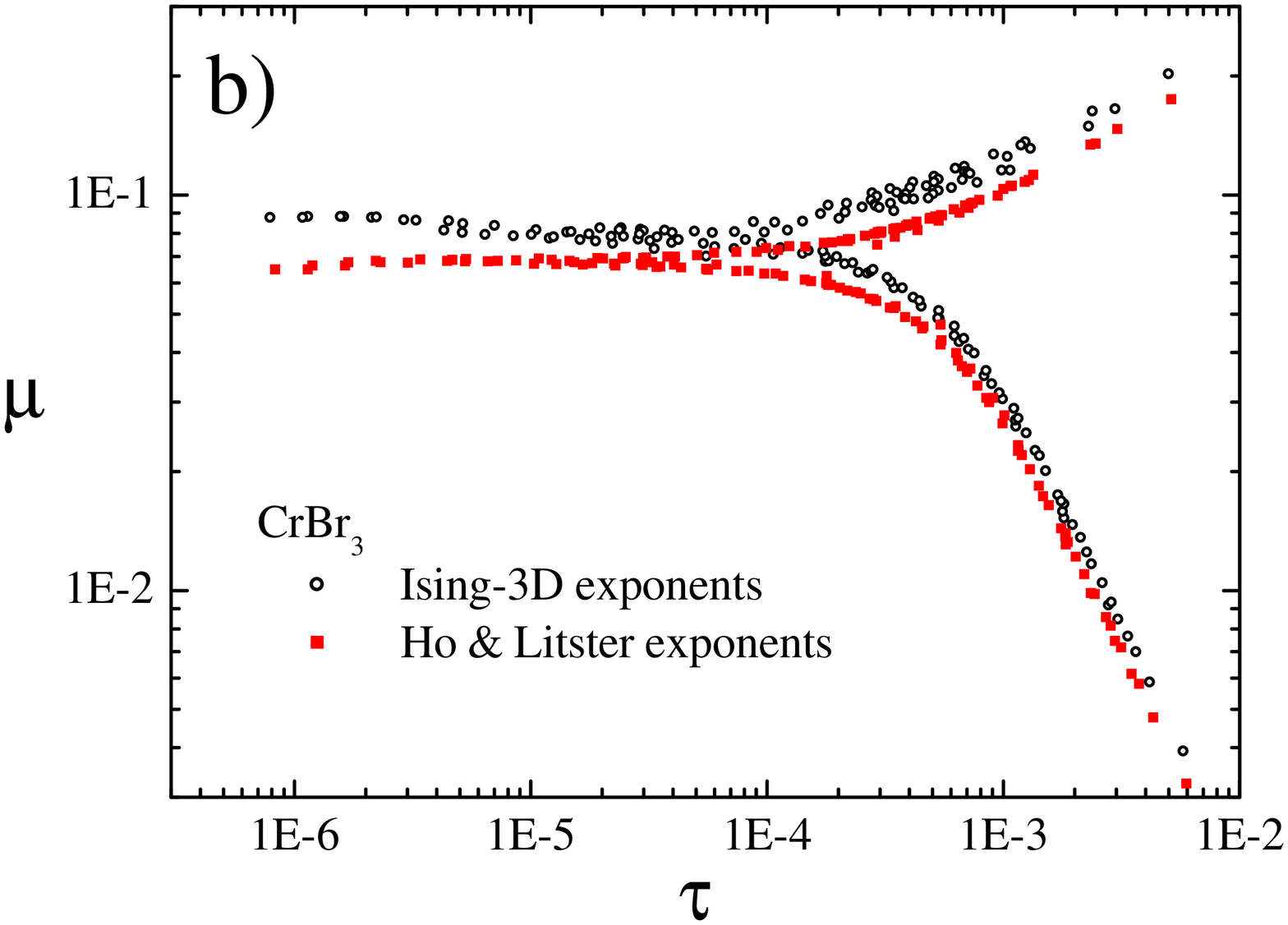}
\caption{\label{fig:epsart3}
Scaling plots of experimental data for $CrBr_3$
(a) $\bar m$$\equiv$$M$/$\mid$$t$$\mid^{\beta}$ vs
$\bar h$$\equiv$$h$/$\mid$$t$$\mid^{\beta\delta}$, and
(b) $\bar \mu$$\equiv$$M$/$h^{1/\delta}$ vs
$\bar \tau$$\equiv$$t$/$h^{1/\beta\delta}$.
The data are made up of 30 isotherms in the interval 
$T_c$-0.9K $< T <$ $T_c$+6.7K. The critical temperature was
$T_c$=32.844 K and the critical exponents used were $\beta$=5/16, $\delta$=5
(Ising 3D) and $\beta$=0.368, $\delta$=4.28, as given in Reference \cite{Ho}.
}
\end{figure}
%
%

Figure 3 gives scaling plots of the high quality data of 30 isotherms at
the vicinity of the Curie temperature pertaining to the insulating ferromagnet
$CrBr_{3}$ measured by Ho and Litster \cite{Ho,Litster} which are a classical example of experimental
scaling data. We show plots of $\bar m$ vs $\bar h$ (Figure 3a) and $\bar \mu$ vs $\bar \tau$
(Figure 3b) with $T=T_{c}$=32.844K for two sets of critical exponents, Ising 3d
(fractional values) and Ho \& Litster (experimental values) summarized in Table \ref{tab:table1}. Both
sets of critical exponents produce good scaling plots in (a) as well as in (b), but
the critical exponents of Ho and Litster produce somewhat better scaling plots.
The accuracy of the magnetization measurements for the set of isotherms was
comparable to that of nuclear magnetic resonance data and it was sufficiently precise
to establish the form of the scaling function. Our Monte Carlo data, shown in
Figures 1(a) and 1(b) are of comparable quality to establish the 3d Ising
scaling function.

%
%
\begin{table}

\caption{\label{tab:table1} Critical exponents}
\begin{ruledtabular}
\begin{tabular}{lccc}

 & $\beta$ & $\delta^{-1}$ & $\gamma$ \\

\hline

Ising 3D \cite{Garcia} & 5/16=0.3125 & 1/5=0.2 & 5/4=1.25 \\

Heisenberg 3D \cite{Yeomans} & 0.340 & 0.208 & 1.39 \\

$CrBr_3$ \cite{Ho} & 0.368 & 0.233 & 1.215 \\

\end{tabular}
\end{ruledtabular}
\end{table}
%
%

%
%
\begin{figure}
\includegraphics[width=6.1cm,height=7.9cm,angle=270]{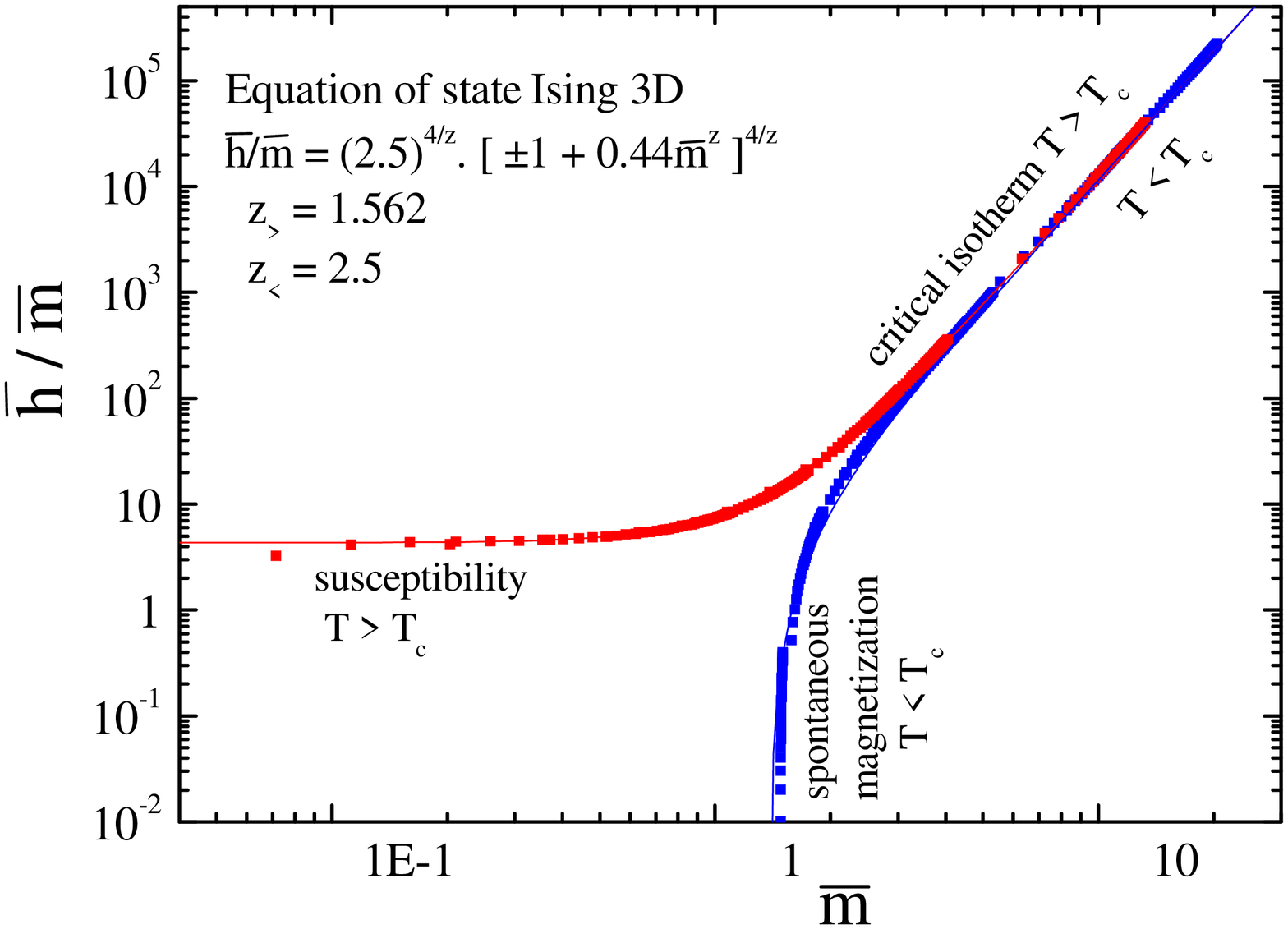}
\caption{\label{fig:epsart4}
Optimized fits of the 30 Monte Carlo isotherms to the given Ising 3D 
equation of state.
}
\end{figure}
%
%

Finally we address the question of the form equation of state for 3D Ising
lattices in the light of the information provided by the set of isotherms in
the vicinity of the critical temperature obtained by the Monte Carlo
method in our $70^3$ s.c. lattice. Figure 4 gives the plot of $M(h,t)$,
Equation (\ref{eq:eq5}), rewritten as

\begin{equation}
\label{eq:eq10}
\frac{\bar h}{\bar m}  = f(\bar m) = A(1 \pm B \bar m^{z})^{(\delta-1)/z}
\end{equation}

Here $A$ can be reduced to unity just by choosing properly the units for the
field $H$. $B$ is a more meaningful coefficient which, in the particular case
of a phase transition describable by means of the mean field approximation
(such as the phase transition in a uniaxial ferroelectric) is equal
to $1/\delta$=1/3. And $z$, as pointed out in Reference \cite{Marques} is
given by $z$$\cong$$\beta\delta/\nu$=2.5 for $T<T_c$ and $z$$\cong$$\beta\delta$=1.562
for $T>T_c$. Figure 4 shows the excellent fit obtained by means of Equation (\ref{eq:eq10})
with ($B/A$)=0.102. The equation of state put in the form given by Equation (\ref{eq:eq10})
is specially good to show directly the most relevant information: (a) the
critical isotherm for $T<T_c$ and $T>T_c$, (b) the spontaneous magnetization curve
($T<T_c$) as a vertical line, and (c) the zero field susceptibility ($T>T_c$) as
a horizontal line. The quality of the fit is comparable or better than those
obtained with traditional expressions of the scaling
function \cite{Ho,Arrot,Vicentini,Schofield,Gaunt,Gonzalo,Milosevic,Milosevic2}.

Work is in progress to obtain Monte Carlo data in larger 3D Ising lattices
at closer field/temperature intervals, and to extend the investigation of
scaling plots in the vicinity of the transition to higher dimensionalities
Ising 4D, Ising 5D, etc, in order to monitor closely how the approach to
mean field behavior takes place. Of course we will be limited to more
reduced sizes (smaller L's) as the dimensionality increases, but we have
excellent experimental data \cite{Jota} on a complete set of isotherms in uniaxial
ferroelectric TGS to produce excellent scaling plots with $T_c$=321.470
and mean field critical exponents $\beta$=1/2 and $\delta$=3.

\begin{acknowledgments}
We specially acknowledge helpful comments and software by M.I. Marqu\'es.
Support from the Spanish MECyD through Grant Number BFM2000-0032 is gratefully acknowledged.
\end{acknowledgments}

\newpage 
\bibliography{apssamp}

\end{document}